\documentclass[pra,twocolumn,showpacs,floatfix]{revtex4}

\usepackage{amsfonts}
\usepackage{graphicx}
\usepackage{dcolumn}
\usepackage{amsmath}
\usepackage{amssymb}
\usepackage{epsfig}

\begin{document}

\title{Encoded Recoupling and Decoupling: An Alternative to Quantum
  Error Correcting Codes, Applied to Trapped Ion Quantum Computation}
\author{D.A. Lidar and L.-A. Wu}
\affiliation{Chemical Physics Theory Group, Chemistry Department,
  University of Toronto, 80 St. George St., Toronto, Ontario M5S 3H6, Canada }

\begin{abstract}
A recently developed
theory for eliminating decoherence and design constraints in quantum
computers, ``encoded recoupling and decoupling'', is shown to be fully
compatible with a promising proposal for an architecture enabling scalable ion-trap quantum
computation [D. Kielpinski \textit{et al.}, Nature \textbf{417}, 709
(2002)]. Logical qubits are
encoded into pairs of ions. Logic gates are implemented using the
S$\o$rensen-M$\o$lmer (SM) scheme applied to pairs of ions at a
time. The encoding offers continuous protection against collective
dephasing. Decoupling pulses, that are also implemented using the SM
scheme directly to the encoded qubits, are capable of further reducing
various other sources of qubit decoherence, such as due to
differential dephasing and due to decohered vibrational modes. The
feasibility of using the relatively slow SM pulses in a decoupling
scheme quenching the latter source of decoherence follows from the
observed $1/f$ spectrum of the vibrational bath.
\end{abstract}

\maketitle

\section{Introduction}

In the quest to construct a scalable quantum computer, a recent proposal 
\cite{Kielpinski:02} advocating an array-based approach to quantum computing
(QC) with trapped ions appears particularly promising. Ions are stored for
later processing in a memory region, then transported to an interaction
region, where pairs are coupled in order to enact quantum logic gates. This
proposal overcomes some of the design constraints that plagued the original
Cirac-Zoller (CZ)\ ion-trap-QC proposal \cite{Cirac:95}, which prevented the
latter from becoming fully scalable. In the new proposal an encoding of a
single logical qubit into the states $\{|0_{L}\rangle =|\!\!\downarrow \uparrow
\rangle ,|1_{L}\rangle =|\!\!\uparrow \downarrow \rangle \}$ of two trapped-ion
(physical) qubits is used. Quantum logic gates are implemented using
the S$\o$rensen-M$\o$lmer (SM) scheme \cite{Sorensen:99,Sorensen:00}
(see also related schemes by Milburn {\it et al.} \cite{Milburn:99,Milburn:00}), which has the
advantage of reduced sensitivity to motional state heating compared to the
CZ proposal. The encoding into $\{|\!\!\downarrow \uparrow \rangle
,|\!\!\uparrow \downarrow \rangle \}$ is useful because these states form a
decoherence-free subspace (DFS) \cite
{Zanardi:97c,Zanardi:97a,Duan:98,Lidar:PRL98} with respect to collective
dephasing, a process whereby the environment introduces identical random
phase modulations on groups of physical qubits \cite{ERD:comment1}. In the
context of the \textquotedblleft quantum charge coupled
device\textquotedblright\ (QCCD) proposed in \cite{Kielpinski:02}, such a
process is one of two dominant sources of decoherence. The DFS encoding
reduces the collective dephasing problem by several orders of
magnitude \cite{Kielpinski:comment}. A
method to perform universal QC using the SM scheme on these DFS qubits was
proposed in \cite{Kielpinski:02}, and independently in \cite{Bacon:thesis}.

The DFS encoding $\{|\!\!\downarrow \uparrow \rangle ,|\!\!\uparrow \downarrow
\rangle \}$ is well known \cite{Palma:96,Duan:98,Lidar:PRL99,Kempe:00},
and its
utility against collective dephasing was demonstrated experimentally using
photons \cite{Kwiat:00} and trapped ions \cite{Kielpinski:01}. The notion of
universal QC using a DFS has been explored in general in \cite{Bacon:99a,Kempe:00,Zanardi:99a,Viola:99a} for Hamiltonians that always
preserve the DFS, in \cite{Lidar:00b,KhodjastehLidar:02,Alber:02}
using a combination of DFS and active quantum error correction methods
\cite{Shor:95,Steane:96a,Aharonov:96,Laflamme:96,Preskill:97a,Knill:98,Steane:02}, and in \cite{Beige:00}, using an approach wherein
transitions out of the code space are suppressed by continuous observation.
Still more generally, the notion of \emph{universal QC while overcoming
decoherence as well as design constraints} has been explored by us and
coworkers at a
theoretical level in a series of recent papers \cite{WuLidar:01,LidarWu:01,WuLidar:01a,WuLidar:01b,ByrdLidar:01a,WuLidar:02,WuByrdLidar:02,LidarWuBlais:02,WuLidar:02a}. This theory uses a combination of qubit encoding into a DFS with
selective recoupling \cite{LidarWu:01} and dynamical decoupling \cite{Viola:98,Viola:98a,Duan:98e,Viola:99,Zanardi:98b,Vitali:99,Vitali:01,ByrdLidar:01,Uchiyama:02} operations, so we refer
to it as \textquotedblleft encoded recoupling and
decoupling\textquotedblright\ (ERD). The utility of ERD as a general method
for quantum simulation, universal QC, and decoherence suppression has also
been stressed and explored by Viola \cite{Viola:01a}.

The main focus of ERD so far was on solid state \cite
{WuLidar:01,LidarWu:01,WuLidar:01a,WuLidar:01b,ByrdLidar:01a,WuLidar:02,WuByrdLidar:02,LidarWuBlais:02,WuLidar:02a}
and NMR \cite{Viola:01a,Fortunato:01} QC proposals. Here we present a \emph{unified treatment} of the ERD ideas, and show that they apply also in an
atomic physics setting, namely the QCCD ion-trap
proposal \cite{Kielpinski:02}. Specifically, we show here how to perform
universal QC on DFS qubits comprised of pairs of trapped ions, by using the
SM scheme for quantum logic, in a manner that involves manipulating only
pairs of ions at any given time, while always perfectly preserving the DFS
encoding (see Brown \textit{et al}. \cite{Brown:02} for an interesting
alternative set of ion pairs-only logic gates, which, however, does not
preserve the DFS at all times). By applying strong and fast dynamical
decoupling (\textquotedblleft bang-bang\textquotedblright\ \cite{Viola:98,Viola:98a})
SM pulses we show how to further drastically reduce sources of decoherence
beyond collective dephasing. While a qubit is being \emph{stored} an example
of such a source of decoherence is deviations from the collective dephasing
approximation. While a qubit is being \emph{manipulated} for the purposes of
QC, coupling to vibrational modes is necessary \cite{Cirac:95,Sorensen:99,Sorensen:00,Milburn:99,Milburn:00}, and decoherence of these vibrational
modes due to patch-potential noise is the second dominant source of qubit
decoherence \cite{Wineland:98,Myatt:00,Kielpinski:01,Turchette:00}. A method
to suppress vibrational mode decoherence (as well as heating, which is not as
serious a problem when the SM scheme is used), employing a version of the
dynamical decoupling method known as \textquotedblleft parity
kicks\textquotedblright , was proposed and discussed in detail by Vitali \&
Tombesi (VT) \cite{Vitali:99,Vitali:01}. This method uses a fast+strong
modulation of the trapping potential. We present here an alternative
decoupling method for suppressing decoherence of ion-trap qubits due to
their coupling to decohered vibrational modes, that operates directly on the
qubit (spin-) states. The feasibility of this scheme, in spite of the
relative slowness of the SM pulses, follows from the observed $1/f$
spectrum of the vibrational bath \cite{Turchette:00,Wineland:comment}. The
concentration of most of the bath spectral density in the vicinity of the
low, rather than the high-frequency cutoff, implies much relaxed
constraints on the decoupling pulses compared to those usually assumed
\cite{Viola:98,Viola:98a,Vitali:99,Vitali:01}. A full analysis of this
result will be presented in a forthcoming publication \cite{ShiokawaLidar:tbp}.

More generally, we show here how \emph{all} sources of
decoherence beyond collective dephasing can in principle be suppressed using
sufficiently strong+fast SM pulses. This includes so-called bath-induced
\textquotedblleft leakage errors\textquotedblright , wherein the system-bath
coupling induces transitions into or out of the qubit subspace \cite{Zanardi:98b,ByrdLidar:01a,WuByrdLidar:02}. We provide feasibility estimates for the
decoupling\ pulses and find that they are within current experimental reach.
The overall picture emerging from this work is that ERD provides a means for
a robust, decoherence-resistant implementation of universal QC with trapped
ions. Experimental implementation of the ERD method should be possible with
current ion-trap technology and we suggest a few experiments.

The structure of the paper is as follows. In section \ref{logic} we review
the DFS encoding into two spins and the associated logic gates. We show how
our previous formulation thereof can be reinterpreted in the context of
acting on pairs of trapped ions within the SM scheme. We also present a
method for coupling pairs of encoded qubits using pulses that involve
controlling only pairs of ions at a time, while always preserving the DFS
encoding. In the subsequent sections we discuss how to reduce decoherence.
In section \ref{decoupling} we review the decoupling method, emphasizing its
application to trapped ion arrays. We then proceed to apply the ERD method:\
in section \ref{createDFS} we show how to eliminate the residual
differential dephasing contribution to decoherence using SM\ pulses; and, in
section \ref{leakage-elim} we discuss how to reduce all further sources of
decoherence, including the component that arises due to coupling to
decohered motional states. Then, in section \ref{all} we show how to fully
implement ERD, i.e., we show how to combine universal QC via recoupling over
DFS-encoded qubits with decoherence suppression via encoded decoupling. To
make ERD fully effective for trapped ions we suggest to combine it with the
VT potential-modulation method. Concluding remarks are presented in section 
\ref{conclusions}.

\section{Encoded universal logic gates in ion traps}

\label{logic}

To fix terminology we first connect the methods developed in \cite
{WuLidar:01,LidarWu:01,WuLidar:01a} to the gates proposed for trapped ions
in \cite{Kielpinski:02}. Let $X_{i},Y_{i},Z_{i}$ denote the standard Pauli
matrices $\sigma _{i}^{x},\sigma _{i}^{y},\sigma _{i}^{z}$, acting on
the $i$th physical qubit (we will use both notations
interchangeably). In \cite{WuLidar:01} it was shown that for the code
$\{|0_{L}\rangle =|\!\!\downarrow \uparrow \rangle ,|1_{L}\rangle =|\!\!\uparrow \downarrow \rangle \}$ the
encoded logical operations (involving the first two physical qubits) are

\begin{eqnarray}
\overline{X}_{12} &=&\frac{1}{2}(X_{1}X_{2}+Y_{1}Y_{2}),\quad  \notag \\
\overline{Y}_{12} &=&\frac{1}{2}(Y_{1}X_{2}-X_{1}Y_{2}),\quad  \notag \\
\overline{Z}_{12} &=&\frac{1}{2}(Z_{1}-Z_{2}).  \label{eq:bars}
\end{eqnarray}
These operations form an $su(2)$ algebra (i.e., we think of them as
Hamiltonians rather than unitary operators). We use a bar to denote logical
operations on the encoded qubits. In \cite{WuLidar:01,LidarWu:01,WuLidar:01a}
these logical operations were denoted by $T_{12}^{\alpha }$, $\alpha \in
\{x,y,z\}$, and a detailed analysis was given on how to use typical
solid-state Hamiltonians (Heisenberg, XXZ, and XY models) to implement
quantum logic operations using this DFS encoding. E.g., the term $
X_{1}X_{2}+Y_{1}Y_{2}$ is the spin-spin interaction in the XY model, and $
Z_{1}-Z_{2}$ represents a Zeeman splitting. A static Zeeman splitting and a
controllable XY interaction can be used to generate a universal set of logic
gates, a result that has very recently been applied in the context of
spin-based QC using semiconductor quantum dots and cavity QED \cite{Feng:02}. Similar conclusions hold when the XY interaction is replaced by a
Heisenberg \cite{LidarWu:01,Levy:01,Benjamin:01} or XXZ interaction \cite{WuLidar:01a}, or even
by a Heisenberg interaction that includes an anistropic spin-orbit
term \cite{WuLidar:02}. We remark that, as first shown in
\cite{Bacon:99a,Kempe:00}, the various types of exchange interactions
can 
be made universal also without any single-qubit terms (such as a Zeeman
splitting), by encoding into three or more qubits \cite{Kempe:00,DiVincenzo:00a,Kempe:01,Kempe:01a,Vala:02}, a result
that has been termed \textquotedblleft encoded
universality\textquotedblright\ \cite{Bacon:Sydney}.

\subsection{Logic gates on two ions encoding a single logical qubit}
\label{SMgates}

S$\o$rensen and M$\o$lmer proposed a quantum logic gate that couples two ions
via a two photon process that virtually populates the excited motional
states of the ions \cite{Sorensen:00}. The SM scheme works well even for
ions in thermal motion, while the CZ scheme requires cooling the ions to
their motional ground state. The SM scheme involves applying two lasers of
opposite detuning $\delta $ to the two ions. Ideally the Lamb-Dicke limit
should be satisfied:

\begin{equation}
(n+1)\eta ^{2}\ll 1,  \label{eq:LD}
\end{equation}
where $\eta $ is the Lamb-Dicke parameter and $n$ is the mean vibrational
number. Deviations from the Lamb-Dicke limit lead to fidelity reduction that
is proportional to $\eta ^{4}$ \cite{Sorensen:00}. The time required to
prepare a maximally entangled state using the SM\ scheme is

\begin{equation}
\tau_{\mathrm{SM}}=\frac{\pi }{\eta \Omega }\sqrt{K}  \label{eq:tauSM}
\end{equation}
where $\Omega $ is the Rabi frequency and $K$ is an integer \cite
{Sorensen:00}. For realistic parameters, in the strong field limit ($K=1$ in
Eq.~(12) of \cite{Sorensen:00}), $\tau _{\mathrm{SM}}$ can be made as short
as $1\mu $sec.

In \cite{Kielpinski:02} it was shown that the SM\ two-ion gate can be
expressed as follows. The unitary gate $U_{2}(\theta ,\phi _{1},\phi _{2})$
was introduced, which we here rename $U_{ij}(\theta ,\phi _{i},\phi _{j})$: 
\begin{eqnarray}
U_{ij}(\theta ,\phi _{i},\phi _{j}) &\equiv &\exp (i\theta X_{\phi
_{i}} X_{\phi _{j}})  \notag \\
&=&\cos \theta I_{i} I_{j}+i\sin \theta X_{\phi _{i}} X_{\phi
_{j}},  \label{eq:Uij}
\end{eqnarray}
where 
\begin{equation*}
X_{\phi }\equiv X\cos \phi +Y\sin \phi .
\end{equation*}
The phase $\phi _{i}$ is the phase of the driving laser at the $i$th ion,
while $\theta \propto \Omega $ and can be set over a wide range of values 
\cite{Sorensen:00,Kielpinski:thesis}. Introducing the operators 
\begin{equation}
\tilde{X}_{ij}\equiv \frac{1}{2}(X_{i}X_{j}-Y_{i}Y_{j}),\quad \tilde{Y}
_{ij}\equiv \frac{1}{2}(Y_{i}X_{j}+X_{i}Y_{j})
\end{equation}
(denoted $R_{ij}^{x}$, $R_{ij}^{y}$ respectively in \cite
{WuLidar:01,LidarWu:01,WuLidar:01a}) we can express

\begin{eqnarray}
U_{ij}(\theta ,\phi _{i},\phi _{j}) &=&\cos \theta \bar{I}+i\sin \theta
(\cos \Delta \phi _{ij}\overline{X}_{ij}  \notag   \\
&&+\sin \Delta \phi _{ij}\overline{Y}_{ij} +\cos \Phi _{ij}\tilde{X}_{ij}+\sin \Phi _{ij}\tilde{Y}_{ij}), \notag   \\
\label{eq:Uij2}
\end{eqnarray}
where $\Phi _{ij}=$ $\phi _{i}+\phi _{j}$. It is simple to check that $
\tilde{X}_{ij}$ and $\tilde{Y}_{ij}$ annihilate the code subspace $
\{|0_{L}\rangle =|\!\!\downarrow \uparrow \rangle ,|1_{L}\rangle =|\!\!\uparrow
\downarrow \rangle \}$ and have non-trivial action (as encoded $X$ and $Y$)
on the orthogonal subspace $\{|\!\!\downarrow \downarrow \rangle ,|\!\!\uparrow
\uparrow \rangle \}\!$. Therefore, as also observed in \cite{Kielpinski:02}
and \cite{Bacon:thesis}, upon restriction to the DFS we can write: 
\begin{eqnarray}
U_{ij}(\theta ,\phi _{i},\phi _{j}) &\overset{\mathrm{DFS}}{\mapsto }&\bar{U}
_{ij}(\theta ,\Delta \phi _{ij})  \notag \\
&=&\exp (i\theta \overline{X}_{\Delta \phi _{ij}})=\cos \theta \bar{I}+i\sin
\theta \overline{X}_{\Delta \phi _{ij}}.  \notag \\
&&  \label{eq:Ubar}
\end{eqnarray}
The fact that $\bar{U}_{ij}$ depends only on the relative phase $\Delta \phi
_{ij}$ is crucial:\ this quantity can be controlled by adjusting the angle
between the driving laser and the interatomic axis, as well as by small
adjustments of the trap voltages (which, in turn, control the trap
oscillation frequency, and hence the ion spacing), whereas it is much harder
to control the absolute phase $\phi _{i}$ \cite
{Sackett:00,Kielpinski:thesis,Kielpinski:02}, and hence also $\Phi _{ij}$.
This is why the code subspace $\{|\!\!\downarrow \uparrow \rangle,|\!\!\uparrow\downarrow \rangle \}$ enjoys a preferred status
over the subspace $\{|\!\!\downarrow \downarrow \rangle ,|\!\!\uparrow
\uparrow \rangle \}$. A thorough 
theoretical analysis of the approximations leading to the gate $
U_{ij}(\theta ,\phi _{i},\phi _{j})$ is given in \cite{Sorensen:00} (see
also \cite{Bacon:thesis} for an abbreviated exposition that
emphasizes the connection to computation in a DFS).

Let us establish the connection between the seemingly distinct sets of logic
operations in Eqs.~(\ref{eq:bars}),(\ref{eq:Uij}). To do so, we only need to
use Eqs.~(\ref{eq:Uij2}),(\ref{eq:Ubar}) while ignoring the component that
annihilates the DFS ($\tilde{X},\tilde{Y}$). Then: 
\begin{eqnarray}
\exp (i\theta \overline{X}_{12}) &=&U_{12}(\theta ,\phi ,\phi )=\bar{U}
_{12}(\theta ,0)  \notag \\
\exp (i\theta \overline{Y}_{12}) &=&U_{12}(\theta ,\phi ,\phi +\frac{\pi }{2}
)=\bar{U}_{12}(\theta ,\frac{\pi }{2})  \notag \\
\exp (i\theta \overline{Z}_{12}) &=&\exp (i\frac{\pi }{4}\overline{Y}
_{12})\exp (i\theta \overline{X}_{12})\exp (-i\frac{\pi }{4}\overline{Y}
_{12})  \notag \\
&=&\bar{U}_{12}(\frac{\pi }{4},\pi /2)\bar{U}_{12}(\theta ,0)\bar{U}_{12}(-
\frac{\pi }{4},\pi /2).  \label{eq:XYZ}
\end{eqnarray}
The third line follows from the elementary operator identity 
\begin{equation}
X_{\phi }=X\cos \phi +Y\sin \phi =e^{-i\phi Z/2}Xe^{i\phi Z/2}
\label{eq:identity}
\end{equation}
which holds for any $su(2)$ angular momentum set $\{X,Y,Z\}$, i.e.,
operators that satisfy the commutation relation $[X,Y]=2iZ$ (and cyclic
permutations thereof), in particular also the encoded operators $\{\overline{
X},\overline{Y},\overline{Z}\}$.

This proves the equivalence of the two sets of operators. Using these
results and Eq.~(\ref{eq:Uij}), a more direct connection can be written in
terms of the Hamiltonians: 
\begin{eqnarray}
\overline{X}_{12}\quad &\Longleftrightarrow &\quad X_{\phi } X_{\phi }
\label{eq:xbar} \\
\overline{Y}_{12}\quad &\Longleftrightarrow &\quad X_{\phi } X_{\phi
+\pi /2},  \label{eq:ybar}
\end{eqnarray}
where the equivalence is meant in terms of a projection of the RHS
Hamiltonians to the DFS. In the context of ion-trap QC the logic gate $\bar{
U
}(\theta ,\Delta \phi )$ can be performed directly, so it may be more
intuitively useful than the $\{\overline{X},\overline{Y},\overline{Z}\}$
set. Eq.~(\ref{eq:XYZ}) shows that by properly choosing $\theta $ and $
\Delta \phi _{ij}$ all single DFS-encoded qubit gates can be performed.

\subsection{Entangling gate between pairs of encoded qubits involving four
ions}

\label{4ions}

In \cite{Kielpinski:02} the following unitary gate was introduced 
\begin{eqnarray}
U_{4} &=&\exp (-i\frac{\pi }{4}X_{\phi _{_{1}}} X_{\phi
_{_{2}}} X_{\phi _{_{3}}} X_{\phi _{_{4}}}) \notag \\
&=&\frac{1}{\sqrt{2}}\left( I_{1} I_{2} I_{3}
I_{4}-iX_{\phi _{_{1}}} X_{\phi _{_{2}}} X_{\phi
_{_{3}}} X_{\phi _{_{4}}}\right)  \notag \\
&&\overset{\mathrm{DFS}}{\mapsto }\frac{1}{\sqrt{2}}\left( \bar{I}
_{12} \bar{I}_{34}-i\overline{X}_{\Delta \phi _{_{12}}} 
\overline{X}_{\Delta \phi _{_{34}}}\right)  \notag \\
&=&\exp (-i\frac{\pi }{4}\overline{X}_{\Delta \phi _{_{12}}} 
\overline{X}_{\Delta \phi _{_{34}}}).
\label{eq:U4}
\end{eqnarray}
This gate, also considered in slightly less general form in \cite{Bacon:thesis}, can be used to entangle two DFS-qubits. It involves
simultaneous control over two phase differences $\Delta \phi _{_{12}},\Delta
\phi _{34}$, and thus control over the motion of two pairs of ions. The case 
$\Delta \phi _{_{12}}=\Delta \phi _{34}=0$ was used in \cite{Sackett:00} to
demonstrate entanglement of four trapped-ion qubits, but this choice is not
unique.

We now come to an important point that was not addressed in \cite{Kielpinski:02}: in order for the DFS encoding to offer protection against
collective dephasing during the exection of the entangling gate, \emph{\
collective dephasing conditions must prevail over all four ions}. To see
this, note that a differential dephasing term such as $(Z_{1}-Z_{3})\otimes
B $ (where $B$ is a bath operator) does not commute with $U_{4}$, so that if
such a term exists during gate execution then the DFS will not be preserved,
according to a theorem in \cite{Bacon:99a}. On the other hand, collective
dephasing over all four ions, expressed by a system-bath coupling of the
form $(\sum_{i=1}^{4}Z_{i})\otimes B$, does commute with $U_{4}$, so that in
this case the DFS is preserved \cite{Bacon:99a}. While deviations from
collective dephasing over pairs of ions have been shown experimentally to be
small \cite{Kielpinski:01}, this may not be the case over the length scales
involving four ions \cite{Kielpinski:comment}. We discuss in section
\ref{createDFS} how to create such extended collective dephasing conditions.

Taken together, the results in this section show how universal QC can be
implemented using trapped ions by applying the SM scheme to pairs of ions at
a time, each encoding a DFS qubit. The DFS encoding takes care of protecting
the encoded information against collective dephasing. We now move on to a
discussion of how to reduce additional source of decoherence.

\section{Dynamical decoupling pulses and their application to trapped ions}

\label{decoupling}

Let us briefly review the decoupling technique, as it pertains to our
problem (for an overview see, e.g., \cite{Viola:01a}). Decoupling, as
proposed by Viola and Lloyd \cite{Viola:98,Viola:98a}, relies on the ability to apply 
\emph{strong and fast} pulses, in a manner which effectively averages the
system-bath interaction Hamiltonian $H_{SB}$ to zero. A quantitative
analysis was first performed in \cite{Viola:98,Viola:98a} for pure dephasing in the
linear spin-boson model (which corresponds to the ohmic case of the
Caldeira-Leggett model \cite{Leggett:87}): $H_{SB}=\gamma \sigma
^{z}\otimes B$, where $B$ is a Hermitian boson operator. The analysis was
recently extended to the non-linear spin-boson model, with similar
conclusions about performance \cite{Uchiyama:02}. Imperfections in the
pulses were considered in \cite{Duan:98e}, and it was shown that an
optimal value for the pulse period can be found. Since the decoupling pulses
are \emph{strong} one ignores the evolution under $H_{SB}$ while the pulses
are on, and since the pulses are \emph{fast} one ignores the evolution
of the bath under its free Hamiltonian $H_B$ during the pulse cycle. The simplest example of eliminating an undesired unitary
evolution $U=\exp [-it(H_{SB}+H_{B})]$, is the ``\emph{parity-kick
  sequence}'' \cite{Viola:98,Viola:98a,Vitali:99}. Suppose we have at our disposal a fully controllable interaction generating a gate $R$ such that
\textquotedblleft $R$ \emph{conjugates} $U$\textquotedblright : $R^{\dagger
}UR=U^{\dagger }$. Then the sequence $UR^{\dagger }UR=I$ serves to eliminate 
$U$. A simple example of a parity kick sequence is the following. Assume we
can turn on the single-qubit Hamiltonian $\Omega X_{j}$ for a time $\pi
/2\Omega $. This generates the single-qubit gate $X_{j}=i\exp (-i\frac{\pi }{
2}X_{j})$. Suppose that $H_{SB}=\sum_{i=1}^{N}\sum_{\alpha \in
\{x,y,z\}}\gamma _{i}^{\alpha }\sigma _{i}^{\alpha }\otimes B_{i}^{\alpha }$. Each term in $H_{SB}$ either commutes or anti-commutes with $X_{j}$. If a
term $A$ in $H_{SB}$ anticommutes with $X_{j}$ then the evolution under it
will be conjugated by the gate $X_{j}$: $X_{j}\exp (-iA\Delta t)X_{j}=\exp
(-iX_{j}AX_{j}\Delta t)=\exp (iA\Delta t)$. This allows for selectively
removing this term using the parity-kick cycle, which we write as:\ $[\Delta
t,X_{j},\Delta t,X_{j}]$. Reading from right to left, this notation means:
apply $X_{j}$ pulse, free evolution for time $\Delta t$, repeat. Suppose
that we can also apply the single qubit gate $Y_{j}$. Then, since every
system factor in the above $H_{SB}$ contains a single-qubit operator, it follows that
we can selectively keep or remove each term in $H_{SB}$ by using the
parity-kick cycle. Note, however, that without additional symmetry
assumptions, this procedure, if used to eliminate \emph{all errors},
requires a number of pulses that is exponential in the number of qubits $N$ 
\cite{Duan:98e,Viola:99}. The reason is that without symmetry assumptions we will
need at least two non-commuting single-qubit operators per qubit (e.g., $X$, 
$Y$), and we will need to concatenate decoupling pulse sequences. Below we
show how to dynamically generate such symmetries, in a way that avoids this
exponential scaling (for a discussion of this point see the Conclusions
section).

Note that in the analysis of the parity kick cycle we ignored $H_{SB}$ and $
H_{B}$ while $R$ was operating; this is justified by the \emph{strength}
assumption. The bath Hamiltonian $H_{B}$ commutes with the applied pulses,
but its effect is very important since $[H_B,H_{SB}]\neq 0$ in
general. Therefore if the bath has spectral components at
frequencies higher than the inverse of the interval between decoupling
pulses, then the bath density matrix will pick up phases that are
essentially random, and this effect will show up as decoherence (for a
quantitative analysis see
\cite{Viola:98,Viola:98a,Duan:98e,Vitali:01,Uchiyama:02}. Hence it is
commonly assumed that the pulse interval, $\Delta t$, should be small compared to the
inverse of the high-frequency cutoff $\omega _{c}$ of the bath spectral
density $I(\omega)$ \cite{Viola:98,Viola:98a}, which also sets the scale of the bath-induced
noise correlation time $t_c$ (the \emph{speed} assumption). It can be shown that
the overall system-bath coupling strength $\gamma _{SB}$ is then
renormalized by a factor $\Delta t\omega _{c}$ after a cycle of decoupling
pulses \cite{Viola:99}, or that the bath-induced error rate is reduced by a factor
proportional to $(\Delta t/t_{c})^2$ \cite{Duan:98e}. Using a Magnus
expansion \cite{Ernst:book}, it can be shown that there
is a hierarchy of decoupling schemes, whereby $\gamma _{SB}$ is renormalized
by a factor $(\Delta t\omega _{c})^{k}$, where $k\geq 1$ is the order of the
decoupling scheme \cite{Viola:99}. The implication for single-qubit
dephasing, $H_{SB}=\frac{1}{2}\gamma _{SB}Z\otimes B$ ($B$ is a
dimensionless bath operator), is that the dephasing time $T_{2}$ is
increased by a factor $1/(\Delta t\omega _{c})^{2k}$
\cite{ERD:comment2}. Thus it seems
crucial to be able to apply pulses at intervals $\Delta t\ll 1/\omega _{c}$.
However, as shown first by Viola \& Lloyd \cite{Viola:98}, and then by VT in their quantitative analysis
of a vibrational mode linearly coupled to a boson bath, a finite-temperature bath sets
another, thermal timescale that must be beat in order for the decoupling
method to work \cite{Vitali:01}. Let the system-bath coupling be 
\begin{equation}
H_{SB}^{\mathrm{vib}}=\gamma \sum_{k}(ab_{k}^{\dagger }+a^{\dagger }b_{k}),
\label{eq:vibHSB}
\end{equation}
where $a$ ($b)$ is an annihilation operator for the system ($k$th bath)
vibrational mode, and $\gamma $ is the (for simplicity uniform) energy
damping rate. In the context of trapped ions the bath is provided by
fluctuating patch-potentials (due, e.g., to randomly oriented domains at the
surface of the electrodes or adsorbed materials on the electrodes) \cite{Turchette:00}. Then VT showed that the decoupling pulse interval (in fact,
the entire cycle time) must be shorter also than the thermal decoherence
time 
\begin{equation*}
t_{\mathrm{dec}}(T)=\{\gamma \lbrack 1+2n(T)]\}^{-1},
\end{equation*}
where $n(T)=[e^{\hbar \omega _{0}/k_{B}T}-1]^{-1}$ is the mean vibrational
number of the system oscillator at thermal equilibrium with temperature $T$,
and $\omega _{0}$ is the frequency of the oscillator, i.e., the system is
described by the harmonic oscillator Hamiltonian $H_{S}=\hbar \omega
_{0}a^{\dagger }a$. Thus the timescale condition for successful decoupling
is
\begin{equation*}
\Delta t\ll \min \{1/\omega _{c},t_{\mathrm{dec}}(T)\}.
\end{equation*}
As shown in the VT analysis, the timescale $t_{\mathrm{dec}}(T)$ is
especially relevant for vibrational mode decoherence in ion traps, which as
already mentioned above, is responsible for qubit decoherence during quantum
logic gate operations.

However, for trapped ions experimental evidence so far points to a $1/f^{\alpha }$ spectrum for the
vibrational bath over a range $1-100$MHz \cite[p.5]{Turchette:00}, implying
that there is no clear high-frequency cutoff $\omega _{c}$. Encouragingly, in a
recent experiment involving a charge qubit in a small superconducting
electrode (Cooper-pair box), a version of parity-kick decoupling was
successfully used to suppress low-frequency energy-level fluctuations
(causing dephasing) due to $1/f$ charge noise \cite{Nakamura:02}. This
suggests that decoupling can help even in the absence of a clear cutoff
frequency. Recent theoretical results support this observation
\cite{ShiokawaLidar:tbp}: for $1/f$ noise most of the bath spectral
density $I(\omega)$ is concentrated in the low, rather than the
high-end of the frequency range.

In spite of the apparent $1/f^{\alpha }$ spectrum in trapped ions, VT used
a cutoff estimate of $\omega _{c}\leq 100$ MHz \cite{Vitali:01}, and showed
that suppression of vibrational decoherence can be accomplished by \emph{pulsing the oscillation frequency }$\omega _{0}$ \emph{of the ion chain}
(i.e., by pulsing the trapping potential), provided $\Delta t<1/\omega
_{c}\sim 1$nsec, \emph{and} $T\leq 10$mK.

Given the estimate in Section \ref{SMgates} of $\tau _{\mathrm{SM}}\gtrsim 1\mu $s for the SM
gate, it is clear that we cannot hope to satisfy the strict $\Delta t<1$nsec
timescale requirement which would be needed in order to use decoupling
directly on the qubit, rather than the vibrational modes, assuming the VT
estimate of $\omega _{c}$. However, the theoretical analysis
\cite{ShiokawaLidar:tbp} and the success of parity-kick decoupling in
the presence of $1/f$ noise in the charge qubit case \cite{Nakamura:02}
suggests that it may well be worthwhile to apply decoupling pulses on the
qubit \emph{in addition} to, or perhaps instead of, the VT
trapping-potential-modulation scheme.

Now let us comment on the strength assumption. Here we must make sure that
the amplitude of the decoupling pulses, $\Omega $, can be made much stronger
than the system-bath interaction $\gamma $ in Eq.~(\ref{eq:vibHSB}), i.e.,
the heating rate from the vibrational ground state of the ion chain.
Experimental measurements of $\gamma $ are very sensitive to trap geometry,
secular frequency, and size \cite{Turchette:00}, and range from a few Hz to
a few tens of KHz \cite{Turchette:00,Roos:99}. On the other hand, one can
have $\Omega $ as high as $1$MHz \cite{Steane:00}, so the strength
assumption can be comfortably satisfied. This does come at a price, however,
since in the strong field limit the SM\ gate is perturbed by a term that
yields direct, off-resonant coupling of the qubit $|\uparrow \rangle $ and $
|\downarrow \rangle $ states without changes in the vibrational motion \cite
{Sorensen:00}. This is a \emph{unitary} gate error that decreases the gate
fidelity by $(N/2)(\Omega /\delta )^{2}$, where $N$ is the number of ions
participating in the gate \cite[Table II]{Sorensen:00}. This forces us to be
in a parameter regime where $\Omega \ll \delta $. In principle it is
possible to exactly cancel this effect if the duration of the laser pulses
is chosen so that both Eq.~(\ref{eq:tauSM})\ and the condition $\tau _{
\mathrm{SM}}=K^{\prime }\pi /\delta $ are satisfied, where $K^{\prime }$ is
an integer and $\delta $ is the detuning. However, in the context of
decoupling we will also need to satisfy conditions such as $\Omega \tau _{
\mathrm{SM}}=\pi /m$ where $m$ is an integer. Putting these conditions
together yields 
\begin{eqnarray*}
\Omega \frac{K^{\prime }\pi }{\delta } &=&\pi /m\quad \Rightarrow \quad
\delta =mK^{\prime }\Omega \\
\Omega \frac{\pi }{\eta \Omega }\sqrt{K} &=&\pi /m\quad \Rightarrow \quad
\eta =m\sqrt{K}
\end{eqnarray*}
While there is no problem with the first of these, the second condition
implies that we cannot be in the Lamb-Dicke limit, Eq.~(\ref{eq:LD}).
Therefore exact cancellation is not possible in our case, and we must resort
to $\Omega \ll \delta $ in order to keep the fidelity reduction to a
minimum. On the other hand, the kind of unitary error that is caused by
off-resonant coupling can be corrected by optimal control pulse shaping
methods \cite{Palao:02}.

Finally, we note that fluctuations of various experimental parameters, such
as intensity and phase fluctuations of the exciting lasers, can cause pure
dephasing of the vibrational modes, in addition to the dissipative coupling
described above \cite{Schneider:98}. Clearly, the success of decoupling
strategies hinges on strong suppression of such fluctuations, as in the
threshold theorem of fault tolerant quantum error correction \cite{Aharonov:96,Preskill:97a,Knill:98,Steane:02}.

To conclude, the discussion in this section indicates that the experimental
viability of decoupling schemes in ion traps is rather promising,
although it is hard to estimate their success at this point. In the
following sections the analysis will be carried out at a more abstract
level, emphasizing the algebraic conditions for a successful implementation
of ERD. In the end it will be up to an experiment to decide the usefulness
of the proposed schemes.

\section{Creating collective dephasing conditions using decoupling pulses:\
reducing decoherence during storage}

\label{createDFS}

One of the important advantages of the DFS encoding $\{|\!\!\downarrow \uparrow
\rangle ,|\!\!\uparrow \downarrow \rangle \}$ is that it is immune to collective
dephasing. However, other sources of decoherence inevitably remain. In this
and the following section, we algebraically classify all additional
decoherence effects and show how they can be eliminated, in particular in
the context of trapped ions.

\subsection{Creating collective dephasing on a pair of ions}

First, let us analyze the effect of breaking the collective dephasing
symmetry, by considering a system-bath interaction of the form 
\begin{equation*}
H_{SB}^{\mathrm{deph}(2)}=Z_{1}\otimes B_{1}^{z}+Z_{2}\otimes B_{2}^{z}
\end{equation*}
where $B_{1}^{z},B_{2}^{z}$ are arbitrary bath operators. This describes a
general dephasing interaction on two qubits, and we can expect this to be
the case during \emph{storage} of trapped ion qubits in the QCCD proposal.
The source of such dephasing during storage is long wavelength, randomly
fluctuating ambient magnetic fields \cite{Kielpinski:01}, that randomly
shift the relative phase between the qubit $|\uparrow \rangle $ and $
|\downarrow \rangle $ states through the Zeeman effect. The interaction can
be rewritten as a sum over a collective dephasing term $Z_{1}+Z_{2}$ and
another, differential dephasing term $Z_{1}-Z_{2}$, that is responsible for
errors on the DFS: 
\begin{equation*}
H_{SB}^{\mathrm{deph}(2)}=\left( Z_{1}+Z_{2}\right) \otimes B_{\mathrm{col}
}^{z}+\left( Z_{1}-Z_{2}\right) \otimes B_{\mathrm{dif}}^{z}.
\end{equation*}
Here $B_{\mathrm{col}}^{z}=\left( B_{1}^{z}+B_{2}^{z}\right) /2$ and $B_{
\mathrm{dif}}^{z}=\left( B_{1}^{z}-B_{2}^{z}\right) /2$. If $B_{\mathrm{dif}
}^{z}$ were zero then there would only be collective dephasing and the DFS
encoding would offer perfect protection \cite{ERD:comment3}. However, in general $B_{\mathrm{dif}}^{z}\neq 0$, and the
DFS encoding will not suffice to offer complete protection.

The crucial observation is that, since $Z_{1}-Z_{2}\propto \overline{Z}_{12}$
[recall Eq.~(\ref{eq:bars})], the offending term causes \emph{logical}
errors on the DFS \cite{ByrdLidar:01a}. Then the problem of $B_{\mathrm{dif}
}^{z}\neq 0$ can be solved using a series of pulses that symmetrize $H_{SB}^{
\mathrm{deph}(2)}$ such that only the collective term remains, as shown in 
\cite{WuLidar:01b,Viola:01a} \cite{ERD:comment4}. To do so note that since the offending term $\propto \overline{
Z}_{12}$, it anticommutes with $\overline{X}_{12}=\frac{1}{2}(X_1 X_2
+ Y_1 Y_2)$. At the same time $
\overline{X}_{12}$ commutes with $Z_{1}+Z_{2}$. This allows us to flip the
sign of the offending term by using a pair of $\pm \pi /2$ pulses in $
\overline{X}_{12}$, while leaving only the collective term. Evolution with
the flipped sign followed by unaltered evolution leads to cancellation of
the offending term. Specifically \cite{WuLidar:01b}:

\begin{equation*}
e^{-iH_{SB}\tau }e^{-i\frac{\pi }{2}\overline{X}_{12}}e^{-iH_{SB}\tau }e^{i 
\frac{\pi }{2}\overline{X}_{12}}=e^{-i(Z_{1}+Z_{2})\otimes B_{\mathrm{col}
}^{z}2\tau },
\end{equation*}
or, in ion-trap terms: 
\begin{equation}
e^{-iH_{SB}\tau }\bar{U}_{12}(-\frac{\pi }{2},0)e^{-iH_{SB}\tau }\bar{U}
_{12}(\frac{\pi }{2},0)=e^{-i(Z_{1}+Z_{2})\otimes B_{\mathrm{col}}^{z}2\tau
},  \label{eq:sym}
\end{equation}
where $\bar{U}_{ij}(\theta ,\Delta \phi _{ij})$ was defined in Eq.~(\ref{eq:Ubar}), and we used the identification found in Eq.~(\ref{eq:XYZ}). This
equation means that the system-bath coupling effectively looks like
collective dephasing at the end of the pulse sequence. Thus, the system is
periodically (every $2\tau $) projected into the DFS.

In order for the the procedure described in Eq.~(\ref{eq:sym}) to work, the
SM\ gate $\bar{U}_{12}(\pm \frac{\pi }{2},0)$ must be executed at a
timescale faster than the cutoff frequency associated with the fluctuating
magnetic fields causing the differential dephasing term in $H_{SB}^{\mathrm{
\ deph}(2)}$. This cutoff has not yet been characterized experimentally, but
the decay rate of the DFS-encoded state of two ions has been measured to be $
2.2$KHz \cite{Kielpinski:01}. Using this as a rough estimate for the cutoff
frequency, we see that the procedure of Eq.~(\ref{eq:sym}) is likely to be
attainable with fast ($\tau _{\mathrm{SM}}\approx 1\mu $s) SM pulses.

\subsection{Creating collective dephasing on a block of four ions}

So far we have discussed creation of collective dephasing conditions on a
single DFS\ qubit. However, as mentioned in Section \ref{4ions}, it is
essential for the reliable execution of an entangling logic gate to have
collective dephasing over all four ions participating in the gate, even if
only two are coupled at a time. A procedure for creating collective \emph{
decoherence} conditions over blocks of $3,4,6$ and $8$ qubits was given in \cite
{WuLidar:01b}. Here we show how to do the same for a block of 4 qubits with
collective dephasing.

Let us start with a general dephasing Hamiltonian on $N$ ions, and rewrite
it in terms of nearest-neighbor sums and differences: 
\begin{eqnarray*}
H_{SB}^{\mathrm{deph}} &=&\sum_{i=1}^{N}Z_{i}\otimes B_{i} \\
&=&\sum_{j=1}^{N/2}\left( Z_{2j}+Z_{2j-1}\right) \otimes B_{2j}^{+}+\left(
Z_{2j}-Z_{2j-1}\right) \otimes B_{2j}^{-},
\end{eqnarray*}
where $B_{2j}^{\pm }\equiv (B_{2j}\pm B_{2j-1})/2$. As noted above, $
Z_{2j}-Z_{2j-1}\propto \overline{Z}_{2j-1,2j}$, so that to eliminate all
nearest-neighbor differences of the form $\left(
Z_{2j}-Z_{2j-1}\right) $ we can use the collective decoupling pulse
$X_{nn}=\bigotimes_{j=1}^{N/2}e^{i\frac{\pi }{2} \overline{X}_{2j-1,2j}}$:

\begin{equation*}
e^{-iH_{SB}\tau }X_{nn}e^{-iH_{SB}\tau }X_{nn}^{\dagger }=e^{-i2\tau
\sum_{j=1}^{N/2}\left( Z_{2j}+Z_{2j-1}\right) \otimes B_{2j}^{+}},
\end{equation*}
or, in ion-trap terms: 
\begin{widetext}
\begin{eqnarray*}
e^{-iH_{SB}\tau }\left[ \bigotimes_{j=1}^{N/2}\bar{U}_{2j-1,2j}(-\frac{\pi }{
2},0)\right] e^{-iH_{SB}\tau }\left[ \bigotimes_{j=1}^{N/2}\bar{U}_{2j-1,2j}(
\frac{\pi }{2},0)\right]
=e^{-i2\tau \sum_{j=1}^{N/2}\left(
Z_{2j}+Z_{2j-1}\right) \otimes B_{2j}^{+}}.
\end{eqnarray*}
The next step is to eliminate next-nearest neighbor differential terms. To
this end let us rewrite the outcome of the $X_{nn}$ pulse in terms of
sums and differences over blocks of four ions:
\begin{eqnarray*}
\sum_{j=1}^{N/2}\left( Z_{2j}+Z_{2j-1}\right) \otimes B_{2j}^{+}
&=& \sum_{j=1}^{N/2}\left[ Z_{2j+2}+Z_{2j+1}+Z_{2j}+Z_{2j-1}\right] \otimes
B_{2j}^{+,+} \notag \\
&+& \sum_{j=1}^{N/2} \left[ (Z_{2j+2}-Z_{2j})+(Z_{2j+1}-Z_{2j-1})\right]
\otimes B_{2j}^{+,-},
\end{eqnarray*}
where $B_{2j}^{+,\pm }\equiv (B_{2j+2}^{+}\pm B_{2j}^{+})/2$. The term in
the first line contains only the desired block-collective dephasing over $4$
ions. The term in the second line contains undesired differential dephasing
terms that we wish to eliminate. But these terms once again have the
appearance of encoded $Z$ operators, between next-nearest neighbor ion
pairs. Therefore we need to apply a second collective pulse $
X_{nnn}=\bigotimes_{j=1}^{N/2}e^{i\frac{\pi }{2}\overline{X}_{2j-1,2j+1}}e^{i
\frac{\pi }{2}\overline{X}_{2j,2j+2}}$, that applies encoded $X$ operators
on these ion pairs. At this point we are left just with collective dephasing
terms on blocks of $4$ ions, as required:
\begin{eqnarray}
e^{-i2\tau \sum_{j=1}^{N/2}\left( Z_{2j}+Z_{2j-1}\right) \otimes
B_{2j}^{+}}\left[ \bigotimes_{j=1}^{N/2}\bar{U}_{2j-1,2j+1}(-\frac{\pi }{2}
,0)\bar{U}_{2j,2j+2}(-\frac{\pi }{2},0)\right] &\times& \notag \\
e^{-i2\tau \sum_{j=1}^{N/2}\left( Z_{2j}+Z_{2j-1}\right) \otimes B_{2j}^{+}}
\left[ \bigotimes_{j=1}^{N/2}\bar{U}_{2j-1,2j+1}(\frac{\pi }{2},0)\bar{U}
_{2j,2j+2}(\frac{\pi }{2},0)\right] &=&
e^{-i4\tau \sum_{j=1}^{N/2}\left(
Z_{2j+2}+Z_{2j+1}+Z_{2j}+Z_{2j-1}\right) \otimes B_{2j}^{+,+}}.\notag \\
\label{eq:create4}
\end{eqnarray}
\end{widetext} This pulse sequence is important to ensure that collective
dephasing conditions will prevail during the execution of logic gates
between DFS\ qubits, as emphasized in Section \ref{4ions}.

To conclude, the procedures discussed in this section provide a means for 
\emph{engineering collective dephasing conditions in an ion trap
experiment}. We propose here to implement these symmetrization schemes
experimentally. Moreover, we propose to combine them with the logic gates
described in Sec.~ \ref{logic}. How to do this efficiently is discussed in
Sec.~\ref{all} below.

\section{Reduction of all remaining decoherence on a single DFS\ qubit
during logic gate execution}

\label{leakage-elim}

The reduction of differential dephasing errors, as in the previous
subsection, is particularly relevant for storage errors. However, this is
only the first step. Additional sources of decoherence may take place during
storage, and in particular during the execution of logic gates, the most
dominant of which is qubit decoherence due to coupling to decohered
vibrational modes, as discussed above. It is useful to provide a complete
algebraic classification of the possible decoherence processes. This will
allow us to see what can be done using SM-decoupling pulses. To this end let
us now write the system-bath Hamiltonian on two physical qubits in the
general form 
\begin{equation*}
H_{SB}=H_{\mathrm{Leak}}+H_{\mathrm{Logi}}+H_{\mathrm{DFS}}
\end{equation*}
where 
\begin{eqnarray}
H_{\mathrm{DFS}} &=&\mathrm{Span}\{\frac{ZI+IZ}{2},\frac{XY+YX}{2},\frac{
XX-YY}{2},  \notag \\
&& ZZ,II\}  \notag \\
H_{\mathrm{Leak}} &=&\mathrm{Span}\{XI,IX,YI,IY,XZ,ZX,YZ,ZY\}  \notag \\
H_{\mathrm{Logi}} &=&\mathrm{Span}\{\bar{X}=\frac{XX+YY}{2},\bar{Y}=\frac{
YX-XY}{2},  \notag \\
\bar{Z} &=&\frac{ZI-IZ}{2}\}  \label{eq:class}
\end{eqnarray}
where $I$ is the identity operator, $XZ\equiv X_{1}Z_{2}$ (etc.), and where $
\mathrm{Span}$ means a linear combination of these operators tensored with
bath operators. The $16$ operators in Eq.~(\ref{eq:class}) form a complete
basis for all $2$-qubit operators. This classification, first introduced in 
\cite{ByrdLidar:01a}, has the following significance. The operators in $H_{ 
\mathrm{DFS}}$ either vanish on the DFS, or are proportional to identity on
it. In either case their effect is to generate an overall phase on the DFS,
so they can be safely ignored from now on. The operators in $H_{\mathrm{Leak}
}$ are the \emph{leakage errors}: terms that cause transitions between
states inside and outside of the DFS. A universal and efficient decoupling
method for eliminating such errors, for arbitrary numbers of (encoded)\
qubits was given in \cite{WuByrdLidar:02}. Finally, the operators in $H_{ 
\mathrm{Logi}}$ have the form of logic gates on the DFS. However, these are
undesired logic operations, since they are coupled to the bath, and thus
cause decoherence.

It is worthwhile to already emphasize that the operator $YI+IY\in H_{\mathrm{
Leak}}$ is of particular importance:\ As shown in \cite[Eq.43]{Sorensen:00},
this is the operator that describes qubit decoherence due to motional
decoherence during application of the SM\ gate.

In the previous subsection we showed how to eliminate the logical error $ 
\bar{Z}$, but we see now that this was only one error in a much larger set.
To deal with the additional errors it is useful at this point to introduce a
more compact notation for the pulse sequences. We denote by $[\tau ]$ a
period of evolution under the free Hamiltonian, i.e., $U(\tau )\equiv \exp
(-iH_{SB}\tau )\equiv \lbrack \tau ]$, and further denote 
\begin{equation*}
P\equiv \bar{U}_{12}(-\frac{\pi }{2},0)=\exp (-i\frac{\pi }{2}\overline{X}
_{12}).
\end{equation*}
Thus Eq.~(\ref{eq:sym}) can be written as: 
\begin{equation*}
\exp [-i(B_{1}^{z}+B_{2}^{z})(Z_{1}+Z_{2})\tau ]=[\tau ,P,\tau ,P^{\dagger
}].
\end{equation*}
Notice that this is an example of a \textquotedblleft
parity-kick\textquotedblright\ pulse sequence.

As a first step in dealing with the additional errors, note that the
symmetrization procedure $[\tau ,P,\tau ,P^{\dagger }]$ can in fact achieve
more than just the elimination of the differential dephasing $Z_{1}-Z_{2}$
term. Since $\overline{X}_{12}$ also anticommutes with $\overline{Y}_{12}= 
\frac{1}{2}(Y_{1}X_{2}-X_{1}Y_{2})\in H_{\mathrm{Logi}}$, if such a term
appears in the system-bath interaction it too will be eliminated using the
same procedure.

So far we have used a $\frac{\pi }{2}\overline{X}_{12}$ pulse.
Interestingly, the Hamiltonian $\overline{X}_{12}$ can also be used to
eliminate all leakage errors \cite{ByrdLidar:01a}. To see this, note that $ 
\bar{U}_{12}(\pm \pi ,0)=\exp (\pm i\pi \overline{X}_{12})=Z_{1}Z_{2}$. This
operator anticommutes with \emph{all} terms in $H_{\mathrm{Leak}}$. Hence it
too can be used in a parity-kick pulse sequence, that will eliminate all the
leakage errors. In particular, this pulse sequence will eliminate qubit
decoherence due to motional decoherence, i.e., the error $YI+IY\in H_{ 
\mathrm{Leak}}$.

At this point we are left with just a single error:\ $\overline{X}
_{12}\otimes B$ itself, in $H_{\mathrm{Logi}}$. Clearly, we cannot use a
pulse generated by $\overline{X}_{12}$ to eliminate this error. Instead, to
deal with this error we need to introduce one more pulse pair that
anticommutes with $\overline{X}_{12}$, e.g., $\exp (\pm i\frac{\pi }{2} 
\overline{Y}_{12})=\bar{U}_{12}(\pm \frac{\pi }{2},\frac{\pi }{2})$.

Let us now see how to combine all the decoherence elimination pulses into
one efficient sequence. First we introduce the abbreviations 
\begin{eqnarray}
\Pi &\equiv &\bar{U}_{12}(\pm \pi ,0)=\exp (\pm i\pi \overline{X}_{12})=\Pi
^{\dagger }=PP  \notag \\
Q &\equiv &\bar{U}_{12}(-\frac{\pi }{2},\frac{\pi }{2})=\exp (-i\frac{\pi }{
2 }\overline{Y}_{12})  \notag \\
\Lambda &\equiv &\bar{U}_{12}(\pm \pi ,\frac{\pi }{2})=\exp (\pm i\pi 
\overline{Y}_{12})=\Lambda ^{\dagger }=QQ  \label{eq:PQetc}
\end{eqnarray}
As argued above, the $\pi $ pulse $\Pi $ eliminates $H_{\mathrm{Leak}}$: 
\[
\exp [-i(H_{\mathrm{Logi}}+H_{\mathrm{DFS}})2\tau ]=[\tau ,\Pi ,\tau ,\Pi ].
\]
\emph{This may be sufficient in practice}, since as argued above this
pulse sequence 
eliminates the $YI+IY$ term, and the DFS encoding takes care of collective
dephasing. Thus we expect that using cycles of two pulses we can almost entirely
eliminate the two most important sources of decoherence. This expectation of
course depends on the time scale requirement for decoupling being satisfied,
as discussed in detail in Section \ref{decoupling} above. In practice it may
well be advantageous to combine the DFS encoding and $[\tau ,\Pi ,\tau
  ,\Pi]$ pulse sequence with the VT method of pulsing the trapping
potential \cite{Vitali:99,Vitali:01}.

Now let us discuss adding the extra pulses needed to achieve full
decoherence elimination. The $\pi /2$ pulse $P$ eliminates $\bar{Y}$ and $ 
\bar{Z}$ in $H_{\mathrm{Logi}}$. Combining this with the sequence for
leakage elimination we have the sequence of $4$ pulses: 
\begin{eqnarray}
e^{-i(H_{\mathrm{DFS}}+\bar{X}\otimes B_{\bar{X}})4\tau } &=&[U(\tau )\Pi
U(\tau )\Pi ]P^{\dagger }[U(\tau )\Pi U(\tau )\Pi ]P  \notag \\
&=&[\tau ,\Pi ,\tau ,P,\tau ,\Pi ,\tau ,P^{\dagger }],  \label{eq:4pulses}
\end{eqnarray}
(where we have used $\Pi P^{\dagger }=P$, $\Pi P=P^{\dagger }$).

If we wish to entirely eliminate decoherence then we are left just with
getting rid of the logical error due to $\overline{X}$. To eliminate it we
now combine with the $\bar{Y}$-direction, $\pi /2$ pulse, $Q$:

\begin{widetext}
\begin{eqnarray}
e^{-iH_{\mathrm{DFS}}8\tau } &=&[U(\tau )\Pi U(\tau )PU(\tau )\Pi U(\tau
)P^{\dagger }]Q^{\dagger }
[U(\tau )\Pi U(\tau )PU(\tau )\Pi U(\tau )P^{\dagger }]Q \notag \\
&=&[\tau ,\Pi ,\tau ,P,\tau ,\Pi ,\tau ,P^{\dagger },Q^{\dagger },
\tau ,\Pi ,\tau ,P,\tau ,\Pi ,\tau ,P^{\dagger },Q]
\end{eqnarray}
\label{eq:10pul}
\end{widetext}
which takes ten pulses. Unfortunately it is not possible to compress this
further, since $P^{\dagger }Q=(i\overline{X})(-i\overline{Y})=i\overline{Z}$
and $P^{\dagger }Q^{\dagger }=-i\overline{Z}$, neither of which cannot be
generated directly (in one step) from the available gate $\bar{U}
_{ij}(\theta ,\Delta \phi _{ij})=\cos \theta \bar{I}+i\sin \theta \overline{
X }_{\Delta \phi _{ij}}$. Finally, note that in principle the last pulse
sequence is applicable also to other QC
proposals, such as NMR and quantum dots.

One important caveat (mentioned in Section \ref{decoupling} above) is that,
because we need very strong and fast pulses, our gate operation may become
imperfect. Specifically, off-resonant coupling and deviations from the
Lamb-Dicke approximation may become important. The former introduces a term $
XI+IX$ into the Hamiltonian generating the $U_{ij}(\theta ,\phi _{i},\phi
_{j})$ gate \cite[Sec. IIIA]{Sorensen:00}, which can cause \emph{unitary}
leakage errors from the DFS. These can in turn be reduced using the methods
in \cite{Tian:00,Palao:02}. Whether the decoupling method we have proposed
offers an improvement will have to be put to an experimental test.

\section{Combining logic gates with decoupling pulses}

\label{all}

So far we have discussed computation using the encoded recoupling method
(Section \ref{logic}), and encoded decoupling (Sections \ref{createDFS},\ref
{leakage-elim}). We now put the two together in order to obtain the full
ERD\ picture. At least two methods are available for combining quantum
computing operations with the sequences of decoupling pulses we have
presented above. For a general analysis of this issue see \cite{Viola:99a}.

\subsection{Fast + Strong Gates Method}

The decoupling pulse sequences given in Sec.~\ref{leakage-elim}
\textquotedblleft stroboscopically\textquotedblright\ create collective
dephasing conditions at the conclusion of each cycle. As noted above, this
is equivalent to a periodic projection into the DFS. This property allows
for \textquotedblleft stroboscopic\textquotedblright\ quantum computation at
the corresponding projection times \cite{Viola:99a}. Here the computation
pulses need to be synchronized with the decoupling pulses, and inserted at
the end of each cycle. The amount of time available for implementation of a
logic gate is no more than the bath correlation time $\tau _{c}=2\pi
/\omega_{c}$. Assuming the dominant decoherence contributions not
accounted for by the DFS encoding to come from differential dephasing
(setting the $\tau_c$ time-scale) and $1/f$ noise, and that we already assumed that we can
use pulses with interval $\Delta t\ll $ $\tau _{c}$, it is consistent to
assume that we can then also perform logic gates on the same time scale.

\subsection{Fast + Weak Gates Method}

There may be an advantage to using fast but weak pulses for the logic gates,
while preserving the fast + strong property of the decoupling pulses. To see
how to combine logic gates with decoupling in this case, let us denote by $
H_{S}=X_{\phi _{i}} X_{\phi _{j}}$ the controllable system
Hamiltonian that generates the entangling gate $U_{ij}(\theta ,\phi
_{i},\phi _{j})$ [recall Eq.~(\ref{eq:Uij})]. Suppose first that we turn on
this logic-gate generating Hamiltonian in a manner that is neither very
strong nor very fast, so that the system-bath interaction is not negligible
while $H_{S}$ is on (this obviously puts less severe demands on experimental
implementation). Then the corresponding unitary operator describing the
dynamics of system plus bath is: 
\begin{equation*}
\tilde{U}(t)=\exp [-it(H_{S}+H_{SB}+H_{B})].
\end{equation*}
Now, \emph{if we choose }$H_{S}$\emph{\ so that it commutes with the
decoupling pulses}, then we can show that after decoupling 
\begin{equation}
\tilde{U}(t)\mapsto \exp [-i2t(H_{S}+H_{B})],  \label{eq:Ucomp}
\end{equation}
provided $t$ is sufficiently small. Tracing out the bath then leaves a
purely unitary, decoherence-free evolution on the system. To prove this,
assume we have chosen $t^{\prime }$ and the decoupling Hamiltonian $
H_{S}^{\prime }$ so that (i) $\exp (-it^{\prime }H_{S}^{\prime })H_{SB}\exp
(it^{\prime }H_{S}^{\prime })=-H_{SB}$ (the parity kick transformation), and
(ii) $[H_{S}^{\prime },H_{S}]=0$. Then 
\begin{widetext}
\begin{eqnarray*}
\tilde{U}(t)e^{-it^{\prime }H_{S}^{\prime }}\tilde{U}(t)e^{-it^{\prime
}H_{S}^{\prime }} &=&\tilde{U}(t)e^{-it[H_{S}+e^{-it^{\prime }H_{S}^{\prime
}}H_{SB}e^{it^{\prime }H_{S}^{\prime }}+H_{B}]} \\
&=&e^{-it(H_{S}+H_{SB}+H_{B})}e^{-it(H_{S}-H_{SB}+H_{B})} \\
&=&e^{-\{2it(H_{S}+H_{B})+t^{2}([H_{SB},H_{S}]+[H_{SB},H_{B}])+O(t^{3})\}},
\end{eqnarray*}
\end{widetext}
where we have used the Baker-Campbell-Hausdorff formula, $\exp (\alpha
A)\exp (\alpha B)=\exp \{\alpha (A+B)+\frac{\alpha ^{2}}{2}[A,B]+O(\alpha
^{3})\}$.

Setting $H_{S}=\Omega S$ and $H_{SB}=\gamma _{SB}S^{\prime }\otimes B$ we
have the condition $t\ll 1/\sqrt{\Omega \gamma _{SB}}$, in order to be able
to neglect the $O(t^{2})$ term $[H_{SB},H_{S}]$. Using $\Omega \sim 1$MHz, $
\gamma _{SB}\sim 10$KHz as in Section \ref{decoupling}, we find $t\ll 10\mu $sec. However, the more stringent constraint comes from the $[H_{SB},H_{B}]$
term, since $H_{B}$ is not bounded for a harmonic oscillator. A\ more
careful analysis then shows the familiar conclusion, that the bath should
not be allowed to evolve for longer than its correlation time \cite{Viola:98,Viola:98a,Vitali:01}. Hence the actual requirement may still the far more
stringent condition $t\ll 1/\omega _{c}\leq 1$nsec for the decoupling pulse
interval; see Sec.~\ref{decoupling}. This cannot be satisfied
with SM pulses, but in this case we can resort to the VT potential
modulation method. When we do this in conjunction with SM decoupling pulses
we can be sure that Eq.~(\ref{eq:Ucomp}) is an excellent
approximation. On the other hand, the requirements for a $1/f$ bath
spectral density are far less stringent and may be satisfied even with
SM pulses alone \cite{ShiokawaLidar:tbp}.
Furthermore, for the rotation angle $\theta =\Omega t$ describing the
computation we have $\theta \ll \sqrt{\Omega /\gamma _{SB}}\leq 10$, which
means that there is no restriction on applying large rotations.

Let us now show how to efficiently combine logic operations and decoupling
pulses. For simplicity consider only the case where we can neglect the $\bar{
X}$ error, i.e., our decoupling sequence is the 4-pulse one given in Eq.~(\ref{eq:4pulses}). Suppose we wish to implement a logical $X$ operation,
i.e., $\exp (-i\theta \overline{X}_{12})$. Recall [Eq.~(\ref{eq:xbar})] that
this involves turning on the Hamiltonian $H_{S}^{X}=\Omega _{X}X_{\phi
} X_{\phi }\overset{\mathrm{DFS}}{\mapsto }\Omega _{X}\overline{X}
_{12}$ between two physical qubits. Because the decoupling pulses $P=\exp (-i
\frac{\pi }{2}\overline{X}_{12})$ and $\Pi =\exp (\pm i\pi \overline{X}
_{12}) $ are generated in terms of the same Hamiltonian, they commute with $
H_{S}^{X}$ while eliminating $H_{SB}$ (except for the terms in $H_{SB}$ that
have trivial action on the DFS). Thus the conditions under which Eq.~(\ref
{eq:Ucomp}) were shown to hold are satisfied. This allows us to insert the logic gates into
the four free evolution periods involved in the pulse sequence of Eq.~(\ref
{eq:4pulses}). Thus, the full pulse sequence that combines creation of
collective dephasing conditions with execution of the logic gate is: 
\begin{equation}
e^{-it(\Omega _{X}\overline{X}_{12}+H_{\mathrm{DFS}})}=\tilde{U}(t/4)\Pi 
\tilde{U}(t/4)P\tilde{U}(t/4)\Pi \tilde{U}(t/4)P^{\dagger },  \label{eq:X}
\end{equation}
with $\tilde{U}(t)=\exp [-it(H_{S}^{X}+H_{SB}+H_{B})]$, and which, using the
DFS encoding, is equivalent to the desired $\exp (-i\theta \overline{X}
_{12}) $. This involves 8 control pulses, 4 of which are of the fast+strong
type (those involving $P$ and $\Pi $), and 4 of which must be fast, but need
not be so strong that we can neglect $H_{SB}$.

If we wish to implement logical $Y$ operation, i.e., $\exp (-i\theta 
\overline{Y}_{12})$, then we cannot now use $P$ and $\Pi $, since they
anticommute with $\overline{Y}_{12}$ and will eliminate it. Instead we
should use decoupling pulses generated in terms of $\overline{Y}_{12}$,
which will also have the desired effect of eliminating $H_{\mathrm{Leak}}$,
as well as $\bar{X}$ and $\bar{Z}$ logical errors, while commuting with the $
\overline{Y}$ logic operations (and for this reason can of course not
eliminate $\bar{Y}$ errors). These are just the $Q$ and $\Lambda $ pulses
defined in Eq.~(\ref{eq:PQetc}). In ion trap terms this implies [recall Eq.~(\ref{eq:ybar})] turning on the Hamiltonian $H_{S}^{Y}=\Omega _{Y}X_{\phi
} X_{\phi +\pi /2}\overset{\mathrm{DFS}}{\mapsto }\Omega _{Y}
\overline{Y}_{12}$ between two physical qubits. Thus: 
\begin{equation}
e^{-it(\Omega _{Y}\overline{Y}_{12}+H_{\mathrm{DFS}})}=\tilde{U}(t/4)\Lambda 
\tilde{U}(t/4)Q\tilde{U}(t/4)\Lambda \tilde{U}(t/4)Q^{\dagger },
\label{eq:Y}
\end{equation}
with $\tilde{U}(t)=\exp [-it(H_{S}^{Y}+H_{SB}+H_{B})]$, and which, using the
DFS encoding, is equivalent to the desired $\exp (-i\theta \overline{Y}
_{12}) $.

Finally, to generate single DFS-qubit rotations about an arbitrary axis we
can combine Eqs.~(\ref{eq:X}),(\ref{eq:Y}) according to the Euler angles
construction. Given that Eqs.~(\ref{eq:X}),(\ref{eq:Y}) each take 8 pulses,
the Euler angle method will generate an arbitrary DFS-qubit rotation in at
most 24 pulses.

Concerning gates that entangle two DFS-qubits, the situation is more
involved, since now the next-nearest neighbor pulses in
Eq.~(\ref{eq:create4}), that create the collective dephasing conditions
on four ions, do not all commute with the $U_4$ gate of
Eq.~(\ref{eq:U4}). Therefore here we must resort to the strong + fast
method of the previous subsection, i.e., we need to synchronize the
$U_4$ pulses with the end of the decoupling pulse sequence.

Taken together, the methods described in this section provide an explicit
way to implement universal QC using trapped ions in a manner that offers
protection against all sources of qubit decoherence, using a fast + strong
(or fast + weak) version of the SM scheme, possibly in combination with the VT potential
modulation method.

\section{Discussion and Conclusions}

\label{conclusions}

We have proposed a method of encoded recoupling and decoupling (ERD) for
performing decoherence-protected quantum computation in ion traps. Our
method combines the S$\o$rensen-M$\o$lmer (SM) scheme for quantum logic gates
with an encoding into ion-pair decoherence-free subspaces (each pair
yielding one encoded qubit), and sequences of recoupling and decoupling
pulses. The qubit encoding protects against collective dephasing processes,
while the decoupling pulses symmetrize all other sources of decoherence into
a collective dephasing interaction. The recoupling pulses are used to
implement encoded quantum logic gates, either during or in between the
decoupling pulses. All pulses are generated directly using the SM scheme. We
have provided numerical estimates of the feasibility of our scheme, which
seem quite favorable. In order to achieve full protection against all
decoherence it may be necessary to supplement ERD with the potential modulation
method due to Vitali \& Tombesi, in order to reduce vibrational mode
decoherence. However, it may be worthwhile to test ERD without potential
modulation first, as a significant reduction in decoherence can already be
expected according to the results presented here. This is so because
the vibrational bath has been found experimentally to have a
$1/f^\alpha$ spectral density \cite{Wineland:comment}, and there exists evidence that in such
a case decoupling may be possible under moderate timing contraints
\cite{ShiokawaLidar:tbp}.

As mentioned in Section \ref{decoupling}, the dynamical decoupling method
requires an exponential number of pulses if the most general form of
decoherence is to be suppressed, that can couple arbitrary numbers of qubits
to the environment (total decoherence \cite{Lidar:PRL98}). This exponential
scaling is avoided here by focusing on decoherence elimination inside blocks
of \emph{finite} size (e.g., at most four ions) where arbitrary decoherence
is allowed. However, we have implicitly assumed that there are no
decoherence processes coupling different blocks. This is a reasonable
assumption for trapped ions, where the different blocks can be kept
sufficiently far apart until they need to be brought together in order to
execute inter-block logic gates. When this happens, ERD can still be
efficiently applied on the temporarily larger block.

It may be questioned whether there is any advantage in using ERD compared
to methods of active quantum error correcting codes (QECC). Both ERD and QECC\ are capable of dealing with
arbitrary decoherence processes, and are fully compatible with universal
quantum computation. There are two main advantages to ERD: First, we need
only two ions per qubit, compared to a redundancy of five ions per qubit to
handle all single-qubit errors in QECC \cite{Laflamme:96}. So far
experiments involving trapped ions have used up to four ions \cite
{Sackett:00}, so that this encoding economy is a distinct advantage for
near-term experiments. Second, our method is directly compatible with the SM
scheme for logic gates in ion traps. On the other hand it is not clear how
to directly use SM gates for QECC. These are general features of ERD:
economy of encoding redundancy and use of only the most easily controllable
interactions. On the other, the disadvantage of ERD compared to QECC is that there does
not exist, at this point, a result analogous to the threshold theorem of
fault tolerant quantum error correction. This means that we cannot yet
guarantee full scalability of ERD as a stand-alone method. However, in
principle it is always possible to concatenate ERD with QECC, as done,
e.g., for DFS with QECC in 
\cite{Lidar:PRL99,Lidar:00b,KhodjastehLidar:02,Alber:02}, and then the standard fault tolerance
results apply.

Finally, we note that ERD is a general method, that is not limited to
trapped ions. We hope that the methods proposed here will inspire
experimentalists to implement encoded recoupling and decoupling in the lab,
thus demonstrating the possibility of fully decoherence-protected quantum
computation, in particular using trapped ions.

\begin{acknowledgments}
This material is based on research sponsored by the Defense Advanced
Research Projects Agency under the QuIST program and managed by the Air
Force Research Laboratory (AFOSR), under agreement F49620-01-1-0468. The
U.S. Government is authorized to reproduce and distribute reprints for
Governmental purposes notwithstanding any copyright notation thereon. The
views and conclusions contained herein are those of the authors and should
not be interpreted as necessarily representing the official policies or
endorsements, either expressed or implied, of the Air Force Research
Laboratory or the U.S. Government. D.A.L. further gratefully acknowledges
financial support from PRO, NSERC, and the Connaught Fund. We thank Prof. C. Monroe and Dr. S. Schneider
for very useful discussions.
\end{acknowledgments}


\begin{thebibliography}{78}
\bibitem[{ERD({\natexlab{a}})}]{ERD:comment1}
 \bibinfo{note}{Formally, collective dephasing on $n$
  qubits is defined by writing down the system-bath interaction Hamiltonian as
  $H_{SB}=\left( \sum_{i=1}^{n}\sigma_{i}^{z}\right) \otimes B$, where $\sigma
  _{i}^{z}$ is the Pauli-$z$ matrix acting on the $i$th qubit, and $B$ is an
  arbitrary bath operator.}
\bibitem[{Wineland({\natexlab{c}})}]{Wineland:comment}
 \bibinfo{note}{It is not clear that $1/f$ noise will hold generally:
 in the event that further experimental progress is made it is
 possible that the only noise limitation remaining will be thermal
 electronic (Johnson) noise, with a flat spectrum, as discussed in
 \cite{Wineland:98} (D. Wineland, private communication). We expect
 that the methods proposed here will still be applicable in that case.}
\bibitem[{ERD({\natexlab{d}})}]{ERD:comment4}
 \bibinfo{note}{The case $B_{\mathrm{oth}}^{z}\neq 0$ can
  also be the result of small departures from equal illumination during the SM\
  logic gate \cite {Kielpinski:01}, as in the intensity and phase fluctuations
  of the exciting lasers analyzed in \cite{Schneider:98}. Hence, in order for
  the decoupling scheme to work the lasers used to implement the corresponding
  SM\ pulses must be more stable than the differential dephasing due to
  magnetic field fluctuation.}
\end{thebibliography}

\end{document}